\documentstyle[aps,prl,twocolumn]{revtex}
\input amssym.def
\input amssym
\font\script=eusm10
\font\bbold=msbm10

\newcommand{\BR}{\mbox{\bbold R}}
\newcommand{\BZ}{\mbox{\bbold Z}}
\begin{document}
\title{The Holographic Entropy Bound and Local Quantum Field Theory}
\author{Ulvi Yurtsever}
\address{Quantum Computing Technologies Group, Jet Propulsion Laboratory,
California Institute of Technology
\newline
4800 Oak Grove Drive, Pasadena, California 91109-8099.}
\date{Received: \today}
\maketitle
\begin{abstract}
The maximum entropy that can be stored in a bounded region of space is
in dispute: it goes as volume, implies (non-gravitational) microphysics;
it goes as the surface area, asserts the ``holographic principle." Here
I show how the holographic bound can be derived from elementary
flat-spacetime quantum field theory when the total energy of Fock states
is constrained gravitationally. This energy constraint makes
the Fock space dimension (whose logarithm is the maximum entropy)
finite for both Bosons and Fermions. Despite the elementary nature of my
analysis, it results in an upper limit on entropy in remarkable
agreement with the holographic bound.

~~~~~

\end{abstract}
\parskip 3pt

~~~~~

An outstanding recent puzzle in gravitational physics is
to find a local, microscopic explanation for the
"holographic principle"~\cite{holoprinciple},
which asserts that the maximum entropy that can
be stored inside a bounded region $\mbox{\script R}$
in 3-space must be proportional to the surface
area $A(\mbox{\script R})$ [as opposed to the
volume $V(\mbox{\script R})$] of the region:
\begin{equation}
S_{\max}(\mbox{\script R})
= \frac{k_B}{4} \frac{A(\mbox{\script R})}{{\l_p}^2} \; ,
\end{equation}
where $k_B$ is Boltzmann's constant, and $l_p = \sqrt{\hbar \, G/c^3 }$
is the Planck length. The most compelling conceptual evidence for the
holographic bound comes from black-hole physics and thermodynamics.
If there were a physical system enclosed in $\mbox{\script R}$
whose entropy exceeded $S_{\max}$, it would be possible to violate
the second law in the following way: First, one could dump as much
energy into $\mbox{\script R}$ as necessary to bring it to the threshold
of gravitational collapse. This process can only increase the entropy
contained in $\mbox{\script R}$, making it exceed $S_{\max}$ even
further. One could then tip the system into full gravitational collapse,
leaving nothing but a black hole inside $\mbox{\script R}$.
The resulting event horizon, being contained in $\mbox{\script R}$,
necessarily has surface area no larger than
$A(\mbox{\script R})$. But according to the Bekenstein
formula~\cite{Bekensteinentropy}, the entropy of this
black hole, given by the right hand side of
Eq.\,(1) with $A(\mbox{\script R})$
replaced by the horizon area, cannot exceed
$S_{\max}$. Thus gravitational collapse would appear
to cause a sudden decrease in entropy, violating
the second law of thermodynamics.

The holographic principle presents a puzzle since derivations
based on standard (non-gravitational) micro-physics yield an entropy
bound proportional to the volume $V(\mbox{\script R})$ instead of
the surface area. To discuss this in the
simplest microscopic model, let me choose $\mbox{\script R}$ to be a
standard three-dimensional spacelike cube
of size $L$ in Minkowski space,
and consider a real, massless
(linear) scalar field $\phi$ confined in $\mbox{\script R}$. The Fock
space is built out of the modes of the
field $\phi$, which are the positive frequency solutions
of the scalar wave equation $\square \phi = 0$ that vanish on
$\partial \mbox{\script R}$. These modes are
given (up to normalization) by the solutions $\sin(\vec{k} \cdot \vec{x} -
\omega_{\vec{k}} \, t)$, where $\omega_{\vec{k}} = c |\vec{k}|$, and
the admissible wave vectors $\vec{k}$ are labelled by non-negative
integers $m_x ,\; m_y ,\; m_z$:
$(k_x , k_y , k_z ) = (\pi /L) (m_x , m_y , m_z)$. I will often use
single-letter labels $i,\; j$ etc.\ to denote a composite
multi-index like $(m_x , m_y , m_z)$. Mode counting and summing
various quantities over the modes (and all my computations below
will be of this kind) can
often be simplified via the standard approximation:
\begin{equation}
\sum_{\vec{k}} \longrightarrow \frac{1}{(\pi /L)^3} \int_{P_{+}} d^3 k
\; = \; \frac{1}{c^3 (\pi /L)^3} \frac{4 \pi}{8}
\int d \omega \, \omega^2 \; ,
\end{equation}
where $P_{+}$ denotes the ``all-positive" octant of $\vec{k}$-space
(consisting of positive $k_x ,\; k_y ,\; k_z$), and the last
simplification is available whenever the summed quantity depends only on
the mode frequency $\omega = c |\vec{k}|$. Consider, for
example, the total number of modes, $N$. A natural cutoff at or
near the Planck frequency, $\omega=2 \pi \mu / \tau_p$, makes $N$
finite, where Planck time $\tau_p = l_p /c$, and
$\mu$ is a dimensionless constant of order 1 to be specified by
a complete theory of the Planck regime (according to
naive Planck-scale physics, $\mu = 1$). The total number of modes
\begin{equation}
N = \sum_{i} \, 1 = 
\frac{L^3}{2 \pi^2 c^3} \int_{0}^{2 \pi \mu / \tau_p } \!\! \omega^2 \, d
\omega = \frac{4 \pi \mu^3}{3} \left( \frac{L}{l_p} \right)^3 \; 
\end{equation}
is proportional to the volume $V(\mbox{\script R}) = L^3$.

The Fock space $\mbox{\script H}_F (\mbox{\script R})$
for the theory can be constructed as the Hilbert space spanned
by orthonormal basis elements of the form
\begin{equation}
|\Psi \rangle
= | n_1 ,\; n_2 , \; \cdots \; , \; n_i , \; \cdots \; , n_N \rangle \; \;
, \; \; \; \; \; \; n_j \in \mbox{\bbold N} \; 
\end{equation}
which denotes a state with $n_i$ particles occupying
mode $i$. With Fermi statistics, each $n_i$ is restricted
to the values $n_i =0 , \; 1$, while with Bose statistics
the $n_i$ can be arbitrarily large integers. The entropy associated
to any quantum state of the field
is given by $S=-k_B {\rm Tr}
(\rho \log \rho )$, where $\rho$ is the density matrix of the state.
The state with the largest possible entropy is the maximally mixed
\begin{equation}
\rho_{\max} = \frac{1}{\dim \mbox{\script H}_F (\mbox{\script R})}
\; \mbox{\large \bf 1} \; ,
\end{equation}
the identity operator normalized by the dimension
of the Fock space $\mbox{\script H}_F
(\mbox{\script R})$.
It follows that maximum entropy is proportional
to the log-dimension of $\mbox{\script H}_F (\mbox{\script R})$:
\begin{equation}
S_{\max} = - k_B {\rm Tr}
(\rho_{\max} \log \rho_{\max} ) =
k_B \log \dim \mbox{\script H}_F (\mbox{\script R}) \; .
\end{equation}
The Fock-space
dimension (and hence the maximum entropy)
is infinite for Bosons unless the number of particles in each mode $i$
is constrained by a finite bound. Assuming that the $n_i$ are so constrained,
\begin{equation}
n_i < D \; , \; \; \; \; \; \; \forall i 
\end{equation}
for some fixed integer $D$, the number of states of the form Eq.\,(4) is $D^N$
[$=\dim \mbox{\script H}_F (\mbox{\script R}) \;$], and Eqs.\,(6) and (3)
give
\begin{equation}
S_{\max} = k_B (\log D) \, N = 
\frac{4 \pi \mu^3 \log D}{3} k_B  \frac{V(\mbox{\script R})}
{l_p^3} \; .
\end{equation}
For Fermions (the case $D=2$),
the maximum entropy is proportional to volume. For Bosons, we must
conclude either that the entropy is unbounded, or we must regularize it
with the occupation-number constraints Eq.\,(7) in which case
the bound is again proportional to volume. Even if the constraints
$D$ were allowed to depend on the mode frequency $\omega_i ,\;$ setting
$D_0 \equiv \min \{ D_i \}$ ($D_0 \geqslant 2$)
it is clear that
$\dim {\mbox{\script H}_F} (\mbox{\script R}) \geqslant
{D_0}^N$, and Eqs.\,(6) and (3) imply
\begin{equation}
S_{\max} \geqslant 
\frac{4 \pi \mu^3 \log D_0 }{3} k_B  \frac{V(\mbox{\script R})}
{l_p^3} \; ,
\end{equation}
still in violent disagreement with the holographic bound Eq.\,(1).

I will now introduce an {\it ansatz}, which consists of imposing an upper
bound on the total energy of states (so that the states are
stable against collapse in semiclassical gravity), and proceed to show that the
resulting constrained Fock space has the right dimension consistent with
the holographic principle. Before proceeding to this calculation and
a discussion of the consequences of the ansatz, a few
comments on the general validity of the holographic bound, and why
I will defer dealing with its generalizations: It has been noted by many
authors~\cite{Bousso1,Bousso2,WaldFlannagan}
that the bound Eq.\,(1) cannot possibly hold for arbitrary spacelike
3-volumes $\mbox{\script R}$.
Even in flat, Minkowski spacetime, it is not difficult to find
examples of $\mbox{\script R}$ for which the bound as given by Eq.\,(1)
is violated. In these examples, the region $\mbox{\script R}$ is
contained in a curved spacelike hypersurface instead of a flat $\{ t = {\rm
const.} \}$ slice of Minkowski spacetime, making different parts of
its boundary $\partial \mbox{\script R}$ Lorentz boosted at
different speeds, and making its surface area arbitrarily small. It is
clear that the thermodynamic argument following Eq.\,(1) above breaks
down for such volumes (the area of the black hole after collapse can exceed
the surface area of $\partial \mbox{\script R}$). A covariant
generalization of the holographic bound\cite{Bousso2,Bousso1},
which replaces the entropy
content of the volume $\mbox{\script R}$ with the entropy contained on
the ingoing null congruence emanating from $\partial \mbox{\script
R}$, appears to have more general validity\cite{WaldFlannagan}. While the
microscopic derivation of the holographic bound I present below is likely to
prevail more generally for Minkowski ($\{ t = {\rm
const.} \}$) volumes with sufficiently
``regular" boundary~\cite{ftnote1}, the more interesting question
of how the derivation is relevant to the covariant holographic principle will
be discussed in a forthcoming paper~\cite{longerpaper}.

Neglecting the small Casimir-effect contribution to the vacuum stress-energy,
the regularized total Hamiltonian for the scalar field $\phi$ can be written
in the form $H = \int_{\mbox{\script R}} : \! T_{00} \! : \, d^3 x
= \sum_i \hbar \omega_i {a_i}^{\dagger} a_i$, where ${a_i}^{\dagger} ,\;
a_i$ are the usual creation and annihilation operators for
the mode $i$. The total energy of a Fock state of the form Eq.\,(4) is
\begin{equation}
\langle \Psi | H | \Psi \rangle
= \langle \Psi | \sum_{i} \hbar \omega_i {a_i}^{\dagger} a_i | \Psi \rangle
= \sum_{i} \hbar \omega_i n_i \; .
\end{equation}
\vspace{-0.2in} 

{\noindent}Let me now introduce the ansatz that the Hilbert
space of the theory
contains only those Fock states $|\Psi\rangle$ for which
\begin{equation}
\langle \Psi | H | \Psi \rangle
= \sum_{i} \hbar \omega_i n_i \; < \; E_{\max} \; ,
\end{equation}
\vspace{-0.2in} 

{\noindent}where $E_{\max} \sim (c^4/G) L$ is an upper bound on energy which
ensures that the field $\phi$ is in a stable configuration
against gravitational collapse
according to semiclassical Einstein equations. More precisely:
{\em the Fock space $\mbox{\script H}_F (\mbox{\script R})$
of the theory consists of the linear span of the (finitely many)
states of the form Eq.\,(4) satisfying the constraint Eq.\,(11).}
It is important to note that this ansatz is consistent
with the linear structure of Fock space;
any $|\Psi \rangle \in \mbox{\script H}_F (\mbox{\script R})$
obeys the same energy bound: $\langle \Psi | H | \Psi \rangle 
< E_{\max}$. For if $|\Psi \rangle$
can be written as a linear combination $|\Psi \rangle
= \sum_{\alpha} c_{\alpha} |{\Psi}_{\alpha}\rangle$,
$\sum_{\alpha} |c_{\alpha}|^2 = 1$, of the basis states
$|{\Psi}_{\alpha}\rangle $ satisfying Eq.\,(11), then, since
$|{\Psi}_{\alpha}\rangle$ are eigenstates of the Hamiltonian $H$,
\[
\langle \Psi | H | \Psi \rangle = \sum_{\alpha} |c_{\alpha}|^2
\langle {\Psi}_{\alpha} | H | {\Psi}_{\alpha} \rangle
< \sum_{\alpha} |c_{\alpha}|^2 E_{\max} = E_{\max} \; .
\]
\vspace{-0.2in} 

{\noindent}Introducing the
dimensionless frequencies $\Omega_i$ and the dimensionless energy bound
$B$ via
\begin{equation}
\Omega_i \equiv \tau_p \, \omega_i \; , \; \; \; \; \; \;
B \equiv \frac{\tau_p}{\hbar} \, E_{\max} \; ,
\end{equation}
\vspace{-0.2in}

{\noindent}the ansatz Eq.\,(11) can be rewritten in the form
\begin{equation}
\sum_{i} n_i \, \Omega_i \; < \; B \; .
\end{equation}
\vspace{-0.2in}

{\noindent}The precise value of $B$ will
depend on the details of a self-consistent
semiclassical (or fully quantum) theory of gravity;
nevertheless, I will assume that it does not differ much from
the value predicted by the hoop conjecture~\cite{hoopcon} applied to the
cube $\mbox{\script R}$:
\begin{equation}
B = \eta \, \frac{\sqrt{3}}{4} \; \frac{L}{l_p} \; ,
\end{equation}
where $\eta$ is a dimensionless number of order 1. According to the
classical hoop conjecture, $\eta = 1$.

What is the dimension of the Fock space constrained as in Eq.\,(11)? For
both Bosons and Fermions, the dimension is equal to the combinatorial
quantity
\begin{eqnarray}
\dim \mbox{\script H}_F (\mbox{\script R})
& = & W(B) \equiv \mbox{\# of} \; (n_1,\; n_2,\; \cdots \; , \; n_N )
, \nonumber \\
& & \; n_i \in \mbox{\bbold N} , \; \mbox{such that} \;
\sum_i n_i \Omega_i < B \; , 
\end{eqnarray}
the cardinality of the space of solutions to Eq.\,(13) in non-negative
integer $N$-tuples. With Fermi statistics, the $n_i$ are further constrained by
$n_i \in \{0 ,1 \}$; for Bosons, there are no additional constraints.
The computation of $S_{\max}$ now
reduces to knowing how to count the quantity $W(B)$.

First the computation for Bosons, since Bose statistics clearly leads to
the larger dimension: Notice that the inequality Eq.\,(13) can be
written in the form
\begin{equation}
\vec{n} \cdot \vec{\Omega} < B \; ,
\end{equation}
where the vectors $\vec{n}=( n_1 , \cdots , n_N )$ and $\vec{\Omega}
=( \Omega_1 , \cdots  , \Omega_N )$ live in $N$-dimensional
Euclidean space ${\BR}^N$. Geometrically, the quantity $W(B)$ is the number of
points of the integer lattice ${\BZ}^N$ which are contained in the convex
subset ${\cal P}^N \equiv
\{ \vec{x} \cdot \vec{\Omega} < B ,\;
x_i \geqslant 0 \} $ of ${\BR}^N$. ${\cal P}^N$
is a polyhedral volume in the positive
$2^N$'th sector ($x_i \geqslant 0 $) of ${\BR}^N$,
bounded by the hyperplane $\{ \vec{x} \cdot \vec{\Omega} = B\}$ (see
Fig.\,1 for the geometry of ${\cal P}^N$ for $N=3$). At first
thought, one might be tempted to conclude that $W(B)$ is simply
proportional to the volume of ${\cal P}^N$, since each unit cell of the
integer lattice $\BZ^N$ contains on average 1 lattice points and has unit
volume. It is easy to show that
the volume of a polyhedron ${\cal P}^n$ in $\BR^n$ whose vertices
(the points where its bounding hyperplane intersects the coordinate
axes) are located at $x_i = l_i , \; i=1, 2, \cdots ,n$, is
\begin{equation}
V({\cal P}^n ) = \frac{1}{n!} \; l_1 \, l_2 \, \cdots \, l_n \; .
\end{equation}
For ${\cal P}^N = \{ \vec{x} \cdot \vec{\Omega} < B , \;
x_i \geqslant 0 \}$, these edge lengths $l_i$ are
\begin{equation}
l_i = \frac{B}{\Omega_i} \; .
\end{equation}
Using $\prod_i (B/{\Omega_i}) = \exp \sum_i \log (B/\Omega _i )$,
$V({\cal P}^N )$ can be calculated with the help of Eq.\,(2);
asymptotically ($L \gg l_p$),
\begin{equation}
V({\cal P}^N ) \sim \frac{1}{N!} \exp \left[
\frac{4 \pi \mu^3}{3} \left( \frac{L}{l_p} \right)^3 \log B \right]
\; .
\end{equation}
According to Eqs.\,(14) and (3) and Stirling's formula $\log N! \sim N
\log N - N$, $V({\cal P}^N )$ vanishes exponentially: $V({\cal
P}^N ) \sim \exp[-N \log (N/B)]$ for large $L/l_p \;$; ${\cal P}^N$ does
not even contain a single lattice point of $\BZ^N$ in its
interior! ~ Solutions of Eq.\,(13) are distributed skin-deep on
the polyhedron ${\cal P}^N$; the bulk of the contribution to $W(B)$
comes from points on the boundary of ${\cal P}^N$ (Fig.\,1).
This boundary is comprised of $N$ polyhedra ${\cal P}^{N-1}$ of
dimension $N-1$, ~ each of which
in turn have boundaries made
\begin{figure}[htp]
\hspace{-.6in}
\centerline{
\input epsf
\setlength{\epsfxsize}{3.000in}
\setlength{\epsfysize}{2.318in}
\epsffile{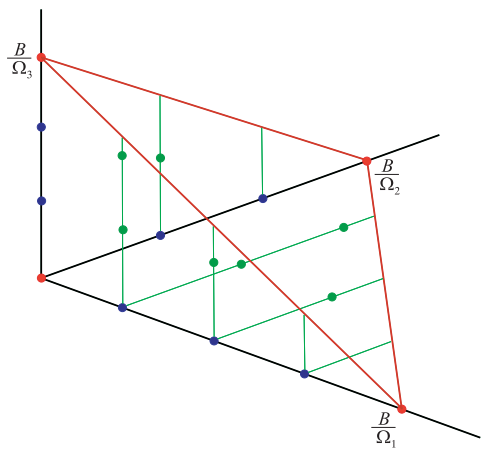}
}
\vspace{0.2in}
\caption[figure]{\label{fig:figure} Example geometry of
${\cal P}^3$, with $\lfloor B/\Omega_1 \rfloor = 4$, $\lfloor B/\Omega_2
\rfloor = \lfloor B/\Omega_3 \rfloor = 3$.
The boundary of ${\cal P}^3$ consists of three ${\cal P}^2$ polyhedra
(the right-triangular walls), three edges (${\cal P}^1$'s), and four vertices.
Contributing to $W(B)$ are 7 points on the ${\cal P}^2$ walls (green dots),
7 points on the edges (blue dots), and four vertices (red dots). There
is only one contributing interior point (not shown); it is located
at $n_1 = n_2 = n_3 =1$.}
\end{figure}
\noindent up of $N-1$ $\; \, {\cal P}^{N-2}$'s, and so on.
By iterating the reasoning above inductively to the lower-dimensional
components of this scaffolding which comprises ${\cal P}^N$'s boundary,
it is not difficult to show that $W(B)$ can be evaluated as
\begin{equation}
W(B) = 1 + N + \sum_{n=1}^{N} \frac{1}{n!} S_n \; ,
\end{equation}

\noindent where, for $1 \leqslant n \leqslant N$,
\begin{equation}
S_n \equiv \sum_{i_1 <i_2 < \cdots < i_n}
(l_{i_1} - 1) (l_{i_2} - 1) \cdots (l_{i_n} - 1) \;  .
\end{equation}
Here I made use of Eq.\,(17) to compute the {\em interior}
volume of each sub-polyhedron ${\cal P}^n$ on the boundary. The edge
lengths $l_{i_k}$ are reduced by 1 so that only interior points
of ${\cal P}^n$ contribute to $W(B)$, and overcounting of points that
lie on the boundaries of each ${\cal P}^n$ is avoided.

Each sum $S_n$ contains $N \choose n$ summands, resulting
in $2^N$ terms in Eq.\,(20). How can Eq.\,(20) be evaluated?
The first key observation is a sequence of elementary algebraic identities
which leads to a recursion relation for $S_n$. If I set
$S_0 \equiv 1$, and introduce the quantities
\begin{equation}
P_m \equiv \sum_{i=1}^{N} (l_i -1)^m \; , \; \; \; \; \; \;
1 \leqslant m \leqslant N \; ,
\end{equation}
then this sequence of algebraic identities are
\begin{eqnarray}
S_0 & = & 1 \; ,  \nonumber \\
S_1 & = & P_1 \; , \nonumber \\
S_2 & = & \frac{1}{2} ( P_1 S_1 - P_2 ) \; , \nonumber \\
S_3 & = &  \frac{1}{3} ( P_1 S_2 - P_2 S_1 + P_3 ) \; , \nonumber \\
\vdots & & \vdots
\end{eqnarray}
leading to the recursion formula
\begin{equation}
S_m = \frac{1}{m} \sum_{j=1}^{m} (-1)^{j-1} P_j \, S_{m-j} \; , \; \; \; \;
1 \leqslant m \leqslant N \; .
\end{equation}
The next key observation is that in
the regime $L/l_p \gg 1$,
\begin{equation}
(P_{1})^m \gg P_m \; , \; \; \; \; \; \; \forall \; m=2,\; \cdots \; , \; N
\; .
\end{equation}
The proof consists of a straightforward evaluation of the sums $P_m$ via
the integral formula Eq.\,(2), which gives
\begin{eqnarray}
P_1 & = & \frac{L^3}{2 \pi^2 c^3} \int_{0}^{2 \pi \mu/\tau_p}
\omega^2 \left(\frac{B}{\tau_p \omega} -1 \right) \, d \omega \nonumber
\\
& = &\mu^2 \left( \frac{L}{l_p} \right)^3 B - N = \mu^2
\left( \frac{L}{l_p} \right)^3 B \left(1 - \frac{4 \pi \mu}{3 B} \right)
\; .
\end{eqnarray}
While for higher $m$ (since lowest $\omega$ is $\pi c/L$,
no true infrared divergences occur at $\omega = 0$),
e.g., for $m \geqslant 4$,
\begin{equation}
P_m \sim \frac{4 \pi}{m-3} \left( \frac{BL}{\pi l_p} \right)^m \; .
\end{equation}
Comparison of Eq.\,(26) with Eq.\,(27) should make Eq.\,(25) obvious
(see~\cite{longerpaper} for full details). It follows that in the
recursion formula Eq.\,(24), the first term of the sum dominates over
all others, proving that asymptotically
\begin{equation}
S_m \sim \frac{1}{m!} \; {P_1 }^m \; ,
\end{equation}
and, by Eq.\,(20) and the asymptotic behavior Eq.\,(19),
\begin{equation}
W(B) = N + q(P_1 ) \; , \; \;
\mbox{where} \; \; q(z) \equiv \sum_{n=0}^{\infty}
\frac{z^n}{(n!)^2} \; .
\end{equation}
To discover the entire analytic function $q(z)$, notice that it
satisfies the differential equation $q_{,zz}+q_{,z}/z -q/z=0$, whose
solutions are Bessel functions of $\sqrt{z}$. Indeed, $q(z)=I_0 (2
\sqrt{z})$, the zeroth-order Bessel function of the second kind~\cite{AS},
with asymptotic behavior as $|z| \rightarrow \infty$:
\begin{equation}
I_0(z) \sim \frac{e^z}{\sqrt{2 \pi z}}
\left[ 1 + \frac{1}{8 \, z} + O\left(\frac{1}{z^{2}} \right)
\right] \; .
\end{equation}
Finally, combining Eqs.\,(29) and (26),
\begin{equation}
W(B) = N + I_0 \left[ 2 \mu 
\left(\frac{B \, L^3}{{l_p}^3} \right)^{1/2} \right] \; ,
\end{equation}
and Eqs.\,(14) and (30) give, asymptotically,
\begin{equation}
S_{\max} =k_B \log W(B) =
k_B \, 3^{1/4} \, \mu \sqrt{\eta} \, \left( \frac{L}{l_p} \right)^2
\; ,
\end{equation}
which~\cite{ftnote2} is in full agreement
with the holographic bound Eq.\,(1) if
$\; \mu \sqrt{\eta} \, = \, 3^{3/4}/2 \, \cong \, 1.14$
[note: $A(\mbox{\script R})=6 L^2$].

With Fermi statistics, the computation of $W(B)$ involves a
different but more straightforward approach, relying on a probabilistic
analysis of the distribution of energy over the $2^N$ subsets (which label the
Fermionic states) of the set of all modes. The result is:
\begin{equation}
S_{\max} = k_B \, \frac{2}{3 \pi \mu}
\, B \, \left[1 + \log\left( \frac{3 \pi \mu}{2} \frac{N}{B} \right)
\right] \; ,
\end{equation}
i.e., $S_{\max}$ is proportional to $(L/l_p ) \log (L/l_p )$. The full
derivation and a discussion of the physical significance of Eq.\,(33)
will be given in~\cite{longerpaper}.

The ansatz Eq.\,(11) does lead to the correct holographic entropy bound, but
how seriously should it be taken? Here are some of the possible
consequences of taking Eq.\,(11) dead seriously as a fundamental
physical law:

The commutation relations (CCR) for Bose statistics
\begin{equation}
[\, a_i \, , \, {a_j}^{\dagger} \, ] = \delta_{ij} \, \mbox{\large \bf 1}
\end{equation}
are incompatible with a finite-dimensional Fock space, as can be
readily seen by taking the trace of both sides of Eq.\,(34) [the result
is: $0= \delta_{ij} \dim \mbox{\script H}_F (\mbox{\script R})$]. Indeed,
according to the ansatz Eq.\,(13), whether they obey the Bose CCR or
the Fermi CAR, the operators ${a_i}^{\dagger}$ must satisfy
the algebraic relations
\begin{equation}
\prod_{i=1}^{N} ({a_i}^{\dagger})^{n_i} = 0 \; \; \;
\mbox{whenever} \; \; \; \sum_{i=1}^{N} n_i \, \Omega_i \geqslant B \; ,
\end{equation}
which imply an algebraic structure drastically different from the CCR
(or CAR). One possible way to specify the new algebra (for Bosons)
is to impose Eq.\,(35) along with
\begin{equation}
[\, a_i \, , \, {a_j}^{\dagger} \, ] = \delta_{ij} \, \mbox{\large \bf 1}
+ C_{ij} \; ,
\end{equation}
where $C_{ij}$ are operators which satisfy
\begin{equation}
{\rm Tr} \, C_{ij} = -\delta_{ij} \,
\dim \mbox{\script H}_F (\mbox{\script R}) = - \delta_{ij} \, W(B) \; 
\end{equation}
and whose matrix elements
$\langle \Psi | C_{ij} | \Psi' \rangle \approx 0$ for
low-energy states $|\Psi \rangle$, $|\Psi' \rangle$. How can this
construction be carried out uniquely, and what are the consequences of
the new algebra for physically observable quantities such as
expectation values of the stress-energy tensor?

An immediate consequence of Eqs.\,(35) and (36) is the breakdown of
Lorentz invariance at scales much earlier than Planck; namely at a new
temperature scale
\begin{equation}
k_B T_c \sim \frac{\hbar c}{\sqrt{L \, l_p}} \; \; .
\end{equation}
For a region $\mbox{\script R}$ of size $\sim \!\! L$, $T_c$ is that
temperature at which massless Bosons confined in $\mbox{\script R}$ have
sufficient thermal energy for gravitational collapse~\cite{ftnote3}.
Relative to the characteristic temperature
$k_B T \!\! \sim \!\! \hbar c/L$, $\; T_c$
corresponds to Lorentz boosts (blueshifts) of order $\gamma \!\! \sim \!\! z
\!\! \sim \!\! \sqrt{L/l_p }$, whereas the Planck temperature
($k_B T_p \!\! \sim \!\! \hbar
c /l_p $) corresponds to (much larger)
boosts of order $\gamma \!\! \sim \!\! z \!\! \sim \!\!
L/l_p $. For feature sizes $L$ at the sub-nucleon scales, the temperature
Eq.\,(38) is reachable via Lorentz boosts that lie only a few orders of
magnitude beyond those envisioned in the large hadron colliders
currently under construction~\cite{bhfactories}.

\vspace{-0.5cm}
\begin{acknowledgements}
\vspace{-0.5cm}
The research described in this paper
was carried out at the Jet Propulsion Laboratory,
California Institute of Technology, under a contract with
the National Aeronautics and
Space Administration (NASA), and was supported by grants
from NASA and the Defense Advanced Research Projects Agency.
\vspace{-0.5cm}
\end{acknowledgements}


\begin{references}
\bibitem{ftnote1} Since the highest-frequency modes make the largest
contribution to the entropy, it appears plausible that the present derivation of
the holographic bound is insensitive to the detailed geometry of the
boundary $\partial \mbox{\script R}$ as long as the average radius
of curvature of $\partial \mbox{\script R}$ is much larger than $\l_p$.
See~\cite{longerpaper} for the calculation with a spherical region
$\mbox{\script R}$, and for an analysis of the more general case.
\bibitem{ftnote2} The result Eq.\,(32) was derived for a Bosonic massless spin-0
field. For higher-spin massless Bosons with two polarization degrees of
freedom (e.g.\ photons), the calculation
is exactly the same~\cite{longerpaper} except the
number of modes is doubled along with the quantity $P_1$, and
$S_{\max}$ is larger by a factor of $\sqrt{2}$. Precise agreement with the
holographic bound Eq.\,(1) then requires
$\; \mu \sqrt{\eta} \, = \, 3^{3/4}/\sqrt{8} \, \cong \, 0.81$.
\bibitem{ftnote3} It is interesting to note that statistical
thermodynamics, in which entropy $\propto L^3 T^3$ (Bose or
Fermi), gives the ``wrong answer" for $S_{\max}$.
See~\cite{longerpaper} for a full discussion.
\end{references}
\end{document}